\documentclass[10pt,letterpaper]{article}
\usepackage[top=0.85in,left=2.75in,footskip=0.75in,marginparwidth=2in]{geometry}

\usepackage[T1]{fontenc}
\usepackage[utf8]{inputenc}
\usepackage{notes2bib}
\usepackage{amsmath}
\usepackage{amsfonts}
\usepackage{nameref,hyperref}

\usepackage[right]{lineno}
\usepackage{microtype}
\DisableLigatures[f]{encoding = *, family = * }
\usepackage[comma,numbers,super,sort&compress]{natbib}
\usepackage{changes}
\usepackage{my_commands}

\usepackage{pdflscape}
\usepackage{longtable}
\usepackage{booktabs}
\usepackage{multirow}

\raggedright
\usepackage{changepage}

\usepackage[aboveskip=1pt,labelfont=bf,labelsep=period,singlelinecheck=off]{caption}

\makeatletter
\renewcommand{\@biblabel}[1]{\quad#1.}
\makeatother

\usepackage{lastpage}
\usepackage{fancyhdr}
\usepackage{graphicx}
\fancyheadoffset[L]{2.25in}
\fancyfootoffset[L]{2.25in}

\usepackage{color}

\definecolor{Gray}{gray}{.25}

\usepackage{graphicx}

\usepackage{sidecap}

\usepackage{wrapfig}
\usepackage[pscoord]{eso-pic}
\usepackage[fulladjust]{marginnote}
\reversemarginpar
\usepackage[comma,numbers,super,sort&compress]{natbib}
\begin{document}
\vspace*{0.35in}

% title goes here:
\begin{flushleft}
{\Large
\textbf\newline{FcsIT: An Open-Source, Cross-Platform Tool for Correlation and Analysis of Fluorescence Correlation Spectroscopy Data
}
}
\newline
% authors go here:
\\
Tomasz Kalwarczyk\textsuperscript{1,*}
\\
\bigskip
\bf{1} Institute of Physical Chemistry, Polish Academy of Sciences\\
Kasprzaka 44/52 01-224, Warsaw, Poland.
\\
\bigskip
* tkalwarczyk@ichf.edu.pl

\end{flushleft}

\section*{Abstract}
% \lipsum[0]
\fcsit\ is a platform-independent, open-source tool for calculating the correlation and fitting fluorescence correlation
spectroscopy data. The software is written in Python and uses a powerful \dpg\ engine for its interface. It provides reading and
correlating the TTTR data, as well as TCSPC filtering of the photon time-trace data. The circular-block bootstrap method applied
to the calculation of correlation data and its variance results in data quality comparable to that obtained with commercially
available software. An intuitive fitting interface provides efficient analysis of large datasets and includes nine predefined
mathematical models for fitting correlation curves. Moreover, it allows users to add their own models in a user-friendly manner.
Validation of the \fcsit\ tool against simulated FCS data and real FCS  experiments confirms its usability and potential appeal to
a wide variety of FCS users.

\section*{Introduction}
Fluorescence correlation spectroscopy (FCS), founded in the 70s of the twentieth century,\cite{Elson1974} has become
one of the standard tools in quantitative biophysics. In the last three decades, the technique and its variations have
been applied to quantify the diffusive propertiesof fluorescent probes and their sizes,\cite{Klapper2015, Shang2017,
Priest2019} monitor biochemical reactions,\cite{Ries2009,
Michalska2018, Kwapiszewska2019} or characterise the properties of the medium in which the tracer
diffuses.\cite{Holyst2009, Kalwarczyk2011,Kwapiszewska2020} The acquisition and analysis of the FCS data require
sophisticated software to handle information from registered photons. There are several software packages capable of
handling the demanding data acquisition and
analysis: \spt\ by PicoQuant, FCSXpert by Becker \& Hickl, or SimFCS. The latter, apart from the analysis of the FCS and
RICS, raster image correlation spectroscopy, also allows the simulation of the FCS experiments. These programs (except
SimFCS) are commercial and are regularly updated. Recently, PicoQuant released UniHarp, a free universal data acquisition
software for PicoQuant systems, enabling the registration and correlation of TTTR data and opening the door to
independent, open-source data analysis software, mostly developed by and for end-users.\\
\indent Across the decades, several free or open-source software projects were developed; some are no longer available or
supported,\cite{Sengupta2003, Krieger2015,
Mueller2014} some are limited to libraries that require programming skills to use,\cite{Ingargiola, Peulen2025, Ries2010}
and others offer limited flexibility and a raw interface.\cite{Waithe2021} The potential of the FCS technique is
continuously growing, and more research groups are adopting it as a standard method in biological and biophysicochemical
studies. As a result, there is an increasing demand for a widely
available and user-friendly tool to analyse FCS data.\\
\indent The \fcsit\ software described in this paper meets these expectations. It is written in Python and uses the
efficient and transparent \dpg\ graphical interface - a modern, fast, and powerful GUI framework. \fcsit\ is a
cross-platform, open-source tool that provides bulk processing of TTTR data, TCSPC filtering, and correlation and
analysis of FCS data. A fast, intuitive interface helps less experienced users better understand the influence of fitting
parameters on the shape of the autocorrelation curve, which is essential for the correct choice of analytical models.
More experienced users will find it beneficial to implement their custom models in a simple text format. The sections
below describe the features and algorithms used in the software; more details are available in the manual deposited on
the repository site.\cite{Fcsit_Kalwarczyk2026}

\section*{Materials and methods}
\subsubsection*{FCS simulations}
\indent Simulations of tracer diffusion were performed using the MCell engine\cite{Stiles1996,Kerr2008} combined with
the FERNET simulator.\cite{Angiolini2015} The MCell input parameters were as follows. The particles of diffusion
coefficient $D=100$
$\mumsq$ were randomly released in a cube with a side of 5.5 $\mum$ at a concentration of 10 nM. The simulation step was
set to 1 $\mus$, and the simulation lasted $10^6$ steps. In a single MCell simulation run, 16 parallel FCS simulations
were performed by running a FERNET simulation in multipoint mode. We set sixteen focal volumes in a 4$\times$4 matrix.
The distance between the focal volumes was set to 0.75 $\mum$, and the dimension of the focal volumes was set to
$\omega_\mathrm{xy}$ = 0.212 $\mum$ and $\omega_\mathrm{z}$ = 1.113 $\mum$, resulting in the structure parameter $\kappa
= 5.25$. The timetraces obtained from FERNET were further correlated and fitted using FcsIT software.

% \subsubsection*{Materials}
% \subsubsection*{Experimental setup}
\subsubsection*{FCS measurements}
\indent FCS experiments using Rhodamine 110 (Sigma-Aldrich) dissolved in water at approximately 20 nM were performed.
Measurements were carried out at 25\textcelsius\ utilising the NIKON C1 confocal setup equipped with the TCSPC
acquisition unit (LSM Upgrade-kit from PicoQuant, Germany). The LSM Upgrade kit consists of a 485 nm pulsed diode laser
controlled by the PDL 828 "SEPIA II" controller, the TCSPC module PicoHarp 300, and a single-photon avalanche diode
(SPAD) detector with appropriate filters. The NIKON PLAN APO water immersion objective x60 NA 1.20 was used. The confocal
unit is equipped with the OKOLab incubator, which provides temperature control with an accuracy of 0.5\textcelsius\. The
data were acquired using the \spt\ software, and the acquisition times were 5, 10, 20, 30, 45, 60, 90, 120, and
150 seconds. For each acquisition time, 5 measurements were carried out. The acquired data were further correlated with
the lifetime correction enabled using the \spt\ software. The obtained correlation curves were exported as ASCII
files and fitted with the \fcsit\ software. The same data, in .ptu format, was also directly loaded into the \fcsit\
software with the TCSPC filtering option enabled. The correlation curves obtained from both programmes were fitted with
the same simple diffusion model and compared.

\section*{Software Description}
\begin{figure*}[ht]
\centering
\includegraphics[width=1\linewidth]{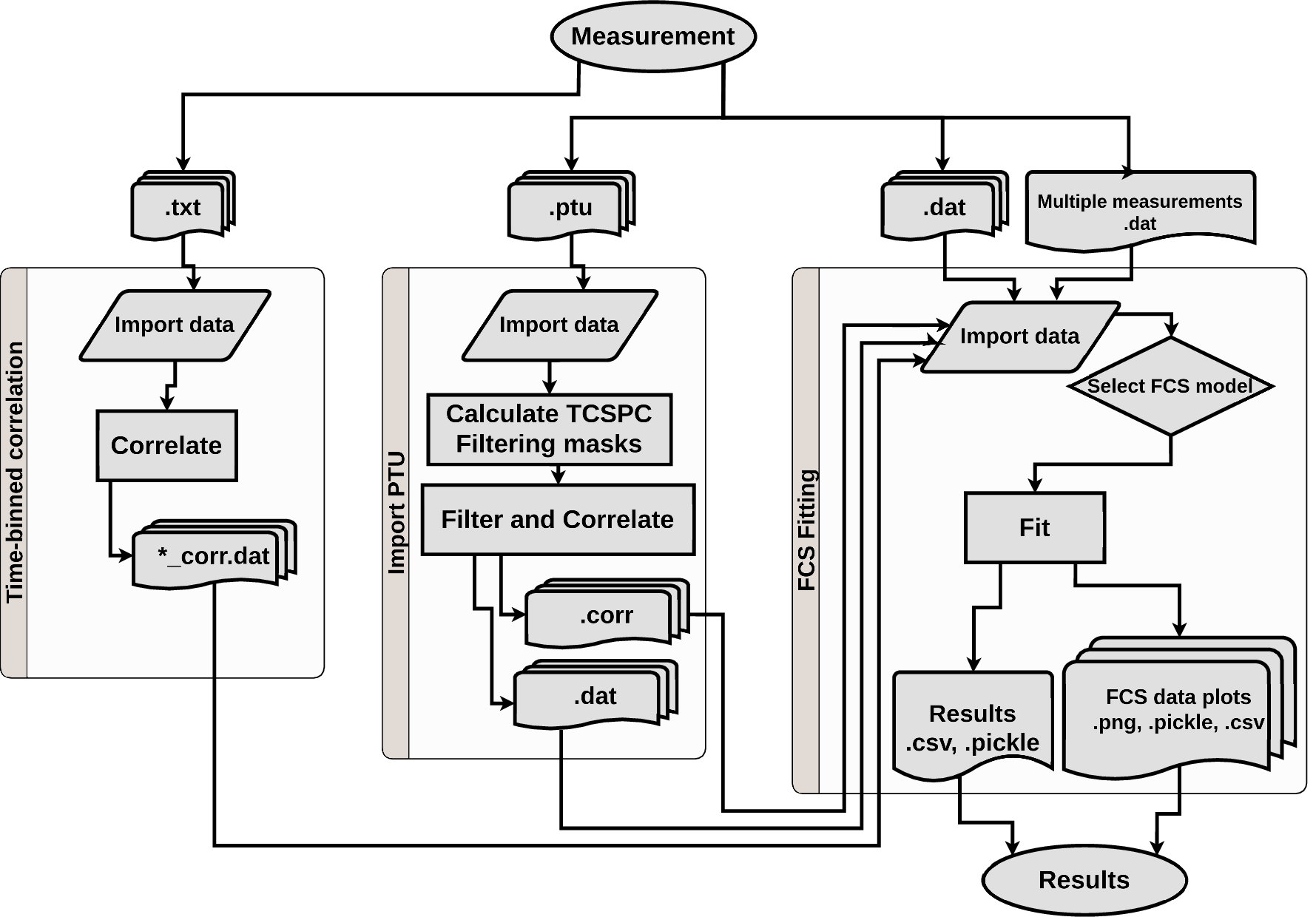}
\caption{\label{fig:flowchart}\textbf{The flowchart that describes the workflow in the \fcsit\ software.}
Depending on their structure and format, experimental data can be analysed in three ways. The simulated binned time-trace data,
stored as \textit{.txt} files, are imported into the \tbcm\ module and then correlated. The correlation curves
are then stored as ASCII \textit{.dat} files, which can be imported directly into the \fitm\ module. The binary data saved by
the \spt\ software, stored in \ptu\ files, is imported into the \ptum\ module. Then, the TCSPC filtration masks are
calculated from the raw TTTR data, and the data are filtered and correlated. The correlated output data can be stored as ASCII
(\textit{.dat}) or binary (\textit{.corr}) files, which, in addition to the correlation data, also include the fitted value of
the
count rate. This feature enables the determination of the molecular brightness of probe molecules. The correlated data can then
be imported into the \fitm\ module. The process of fitting the correlation data is performed in the \fitm\ module. It
can import correlated data directly from \spt\ or calculate it within \fcsit. The output data can be stored as plots, in
various formats, and as a table containing the fitting parameters for each fitted file.}
\end{figure*}

\indent The \fcsit\ software is written as a Python script that uses the \dpg\ graphical interface. The software consists of three
main modules: \textsl{i)} the \tbcm\ the module for correlating the binned timetrace data; for
example from simulations of FCS experiments,
\textsl{ii)} the \ptum\ module for reading, filtering, and correlating the experimentally obtained TTTR data
storred in the binary format. Currently, this option is
limited to \ptu\ files created with the \spt\ data-acquisition software provided by PicoQuant, Germany. \textsl{iii)}
the \fitm\ module
for reading and fitting the correlation data (ASCII files or data correlated using the \fcsit). The scheme of the workflow is
depicted in Figure \ref{fig:flowchart}. For detailed manual, see the repository site.
\cite{Fcsit_Kalwarczyk2026}\\

%
% \subsection*{Correlation of the simulated timetraces}
%
\subsection*{Time-binned correlation module}
\indent This module is designed to read and correlate previously binned fluorescence fluctuations time-trace data, whether
simulated or experimental. The module reads single-column data, where each row takes either 0 (no photons) or $\Nphot$, the number
of photons observed in a given time bin. The data should be binned equally, with each row corresponding to the user-defined time
step.\\
\indent To calculate the autocorrelation curves, the photon time trace is divided into chunks. The lengths of chunks are equal by
default, but the user can also define the chunks' positions and lengths manually.
Next, each chunk is independently correlated using the \emph{multipletau} library,\cite{Mueller2012Multipletau} returning
logarithmically
distributed lag times. The data frame containing correlation data for all chunks is passed to a function that calculates the mean
and variance of the correlation curve using the circular-block bootstrap method described below.\\
\begin{figure}[t!]
\includegraphics[width=1\linewidth]{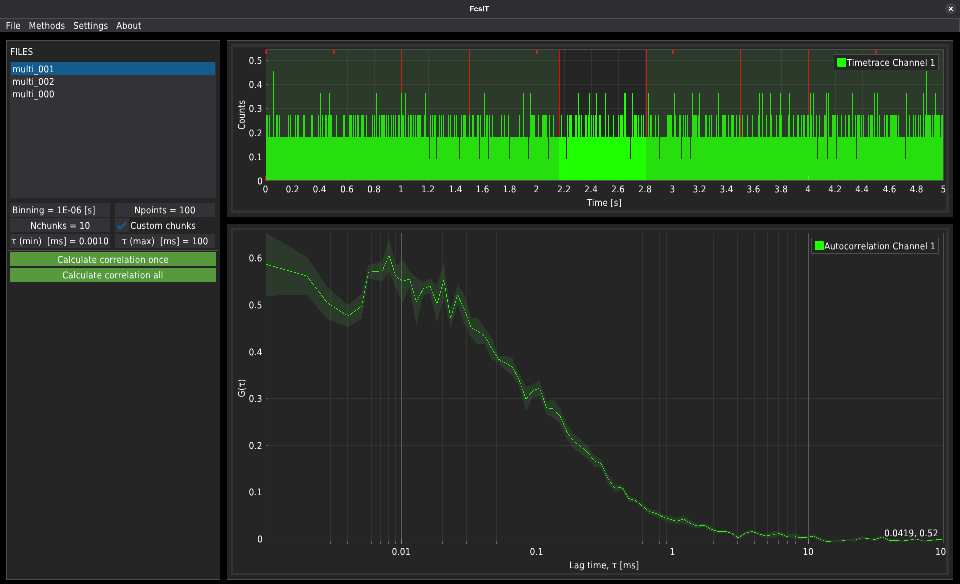}
 \caption{\label{fig:TBC-interface}
 \textbf{The interface of the Time-binned correlation module.}
Starting from the top, the left panel consists of the list of files, the sliders defining the correlation parameters including:
time binning (default 1 $\mus$), Number of points in the resulting autocorrelation curve, number of chunks in which the whole
time trace is divided, custom chunks witch that turns on and off the customizable cunks’ positions and length, and th eminimal
and amximal lag times. The top central-right panel displays the time-trace signal, with the correlation curve below it. The
shaded area on the time-trace corresponds to part of the signal used to calculate the correlation curve. The shaded area on the
correlation plot corresponds to the standard error of the correlation $\Gtau{}{}$ at a given $\tau$. The mean value of the
correlation curve (averaged over chunks) and the corresponding error are calculated according to the procedure described in the
\textit{Calculating the mean and error of correlation curves} section.
}
\end{figure}

\indent Figure \ref{fig:TBC-interface} shows the interface of the Time-binned correlation module. The software has been tested on
data generated from the MCell FERNET FCS simulations.\cite{Stiles1996,Kerr2008,Angiolini2015} The simulated
single-fluorescence-channel data were imported and correlated. The left panel lists files imported from a given directory. Below
are the time-trace and correlation settings, including the time bin, the number of correlation points in the correlation curve,
and the number of chunks. Here, the number of chunks was decreased to 4 for demonstration purposes to visualise the possibility of
changes in chunk length. In the top centre panel, there is a full-time trace. At the same time, the shaded areas are marked with
the red markers corresponding to the signal included in
the calculations of the correlation curve shown in the lower centre panel. Between chunks 2 and 3, there is an intentionally
left-uncorrelated gap.\\

%
% \\
% \input{Tab_timing}
%
%
\subsection*{Import PTU method}
\indent This module was designed to operate on raw experimental data acquired with the commercially available fluorescence
correlation spectroscopy setup from PicoQuant. The \spt\ software from PicoQuant was used for data acquisition.
\spt\ stores the time-tagged time-resolved (TTTR) data in the binary \ptu\ files. The TTTR signal contains information about
the time of photon acquisition, its time delay after the excitation event (fluorescence lifetime), and the channel in which the
detector registered it. Currently, the software is designed to recognise up to two detectors. These features allow filtering of
the experimental data to remove artefacts such as afterpulsing, or to perform fluorescence lifetime correlation spectroscopy
(FLCS). In the current version of the software, only the background/afterpulsing filtration method is available. Still, the
project's open-source nature leaves room for further development.\\
\indent To read the .ptu data files, \fcsit\ uses the modified readPTU\_FLIM library,\cite{Rohilla2019} which, instead of opening
the
fluorescence lifetime imaging (FLIM) data files, reads the FCS files recorded in point mode. If two fluorescence channels are used
(two detectors), the photon time trace is split into two time traces, each corresponding to a separate fluorescence channel. Next,
for each fluorescence channel, the TCSPC histogram is calculated. The software recognises whether the timetrace was acquired in a
standard or pulse-interleaved excitation (PIE) mode and automatically accounts for the shift between the lasers' excitation pulses
in the PIE mode. Based on the TCSPC histogram, the user decides how to filter the TTTR data.\\
\indent Two filter types are available, and each channel can be filtered individually. The first one uses the simplest time
gating
method.\cite{Rich2013} In this method, only those photons are taken into account whose time of acquisition after the excitation
pulse is in the user-defined range. Photons outside the range are removed from the time trace. The second type of time-resolved
filtering method is based on the fluorescence lifetime decay pattern, using a background/afterpulsing removal method. First, the
user determines the background level (typically the plateau part of the TCSPC histogram at long delay times). Next, the
determined
background intensity value is subtracted from the total signal. The as-obtained pattern is further used for calculations of the
weights ascribed to each photon.\\

\begin{figure}[t!]
\includegraphics[width=1\linewidth]{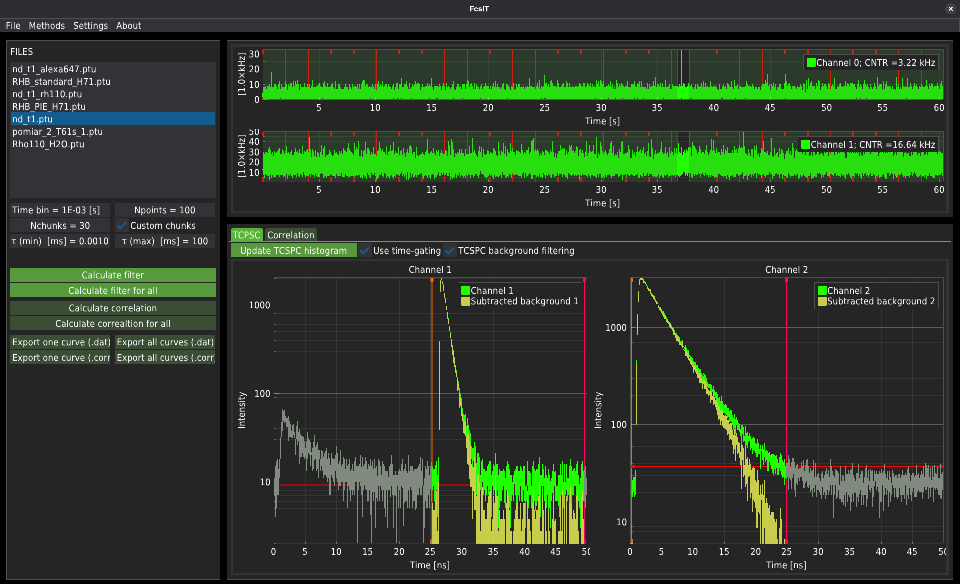}
 \caption{\label{fig:PTUIMP-interface}\textbf{The interface of the PTU Import module.} The left panel is similar to the
one depicted in Figure \ref{fig:TBC-interface}. The main difference is that the time bin influences only the displayed
timetrace, not the calculations. Below the correlation controls, there are buttons to calculate filtering masks and to
calculate the correlation data. Finally, there are export buttons to save the calculated correlation curves. In the top
centre, there are time-trace plots (one for single-channel data or two for two-channel data). Below are either TCSPC
histograms for each channel (when the \textit{TCSPC} tab is active) or correlation curves (when the \textit{Correlation}
tab is active).}
\end{figure}

\indent The weights are calculated according to the algorithm described in references \citenum{Enderlein2005} and
\citenum{Kapusta2007}. Next, the time traces of weighted photons are divided into chunks. Similar to the method described in the
previous section, the chunks are lengthened by default equally, but the user can manually modify their positions and lengths. This
feature may be of interest for measurements in biological systems (for example, in living cells), where large artefacts can be
observed in a time trace. The user can selectively remove unwanted parts of the time trace and correlate only the selected portion
of the signal. For each chunk, the auto- and/or crosscorrelation curves are calculated separately. The correlation data for all
chunks is then passed to the function that calculates the mean and variance of the correlation data
using the circular-block bootstrap method described in the next section.\\
\indent To correlate the timetrace data, the algorithm described by Wahl et al.\cite{Wahl2003} is used. This algorithm was later
translated into the Python library by Dominic Waithe.\cite{Waithe2021} FcsIT uses the modified version of D. Waithe’s original
tttr2xfcs function. The main modifications are: \textit{i)} added definitions of the lag-time range over which the correlation is
calculated ($\tau_\mathrm{min}$ and $\tau_\mathrm{max}$), \textit{ii)} vectorised calculations, \textit{iii)} using pure Python
instead of Cython to make FcsIT more platform-independent.\\

\subsection*{Calculating the mean and error of correlation curves}

\indent Knowledge of the correct error values at each point along the correlation curve is necessary for accurate evaluation of
the data and the fit of the theoretical model to the experimental data. The classic approach to estimating correlation errors was
described by Wohland et al.\cite{Wohland2001} Briefly, they compared several error-estimation methods with the approach proposed
by Koppel.\cite{Koppel1974} The models included: \textit{i)} errors as the standard error of mean calculated for each lag-time
point, \textit{ii)} errors equal to standard deviation (SD) obtained by averaging of multiple correlation curves, and finally
\textit{iii)} the errors as the $\sigmean=\mathrm{SD}\slash\sqrt{M}$, where $M$ stands for the number of chunks into which the
whole time trace is divided. The latter model was considered most preferable when a single correlation curve is considered. It
assumes, however, that all $M$ chunks are independent, there are no correlations between chunks, and all chunks are equally
weighted. These assumptions fail when the chunks are correlated (for example, due to drift or oscillations in the laser power) or
when the number of photon pairs characterised by the same lag times $\tau$ significantly differs between chunks (due to variation
in chunk length). \\
\indent In the FCS data, the correlation curves $\Gtau{1}{},...,\Gtau{M}{}$ are calculated for chunks that are ordered along the
time axis of the time trace. It means that there is a non-zero probability that the neighbouring chunks are not independent. To
address this, commercially available software uses bootstrap methods that randomly analyse multiple parts of the signal to
estimate errors without knowledge of their distribution. Nevertheless, the detailed information on the exact algorithms used by
the software developers is limited. FcsIT implements the circular-block bootstrap method, which is typically used for its ability
to preserve block correlations.\cite{Politis1991,Haerdle2003,Kondo2025} The algorithm is described below.\\
\indent For each of $M$ chunks that can vary in length, the correlation curve $\Gtau{j}{}$ for $j=1,...,M$ is calculated. As a
result, a matrix $X_{i,j}=\Gtau{j}{i}$ is obtained where $i=1,...,T_n$ is the index of the lag-time vector, and
$X\in\mathbb{R}^{T_n\times M}$. For each $\tau_i$, an effective bin width $\Delta\tau_i$ is define such that
\begin{equation}
\label{eq:binwidth}
\Delta \tau_i =
\begin{cases}
\tau_{2} - \tau_{1}, & i = 1,\\
\tau_{i+1} - \tau_{i-1}, & 1 < i < T_n,\\
\tau_{T_n} - \tau_{T_n-1}, & i = T_n.
\end{cases}
\end{equation}
Next, for each lag-time value and for each chunk, the number pairs of photons $K_{i,j}$ contributing to a certain
$\Gtau{j}{i}$ value is calculated:
\begin{equation}
\label{eq:Kij}
K_{i,j} =
\begin{cases}
\lfloor\frac{T_j-\tau_i}{\Delta\tau_i}\rfloor\\
0, & \lfloor\frac{T_j-\tau_i}{\Delta\tau_i}\rfloor<0
\end{cases}
\end{equation}
To estimate the mean correlation curve $\Gbar$
and its variance, the $B$ circular-block bootstrap replicates over $\nb$
overlapping blocks is performed. $\nb$ is calculated from the number of chunks and the length of the blocks $L=3$ as $\nb =\lceil
M\slash
L\rceil$. For each replicate $b = 1,...,B$, the vector of bootstrap resamples of chunks $J_{b,j}$ is then calculated. The
bootstrap mean
for the $b$-th replication, weighted over the number of pairs $K_{i,j}$ is defined as:
\begin{equation}
 \label{eq:bootstrap_rep_mean}
\Gtilde = \frac{\sum_{j=1}^MK_{i,j}X_{i,J_{b,j}}}{\sum_{j=1}^MK_{i,j}}
\end{equation}
The bootstrap estimator of the mean correlation value at $\tau_i$ equals to:
\begin{equation}
 \label{eq:meanG}
 \Ghat=\frac{1}{B}\sum_{b=1}^{B}\Gtilde,
\end{equation}
and its variance is defined as:
\begin{equation}
 \label{eq:VarG}
 Var\left(\Ghat\right)=\frac{1}{B-1}\sum_{b=1}^B\left(\Gtilde-\Ghat \right)^2.
\end{equation}
\indent In the circular block bootstrap method, the variance is biased downward and needs to be corrected by the factor equal to
$M\slash\nb^{\mathrm{eff}}$, where $\nb^{\mathrm{eff}} = M-L+1$.\cite{Davison1997} The standard error $\mathrm{SE}$ of the mean
correlation curve $\Ghat$ is therefore calculated as:
\begin{equation}
 \label{eq:SE}
\mathrm{SE}  =\sqrt{Var\left(\Ghat\right)\frac{M}{\nb^{\mathrm{eff}}}}
\end{equation}
\indent To maintain numerical stability, several corrections are applied. For example, to calculate $\Gtilde$ in equation
\eqref{eq:bootstrap_rep_mean}, the values for which $\sum_{j=1}^MK_{i,j}=0$ are excluded from the analysis. Also, in
real data, missing or undefined correlation values may occur for some chunks and lag times, especially when the chunks
differ in length. In such cases, only the finite values of the weighted averages are included in calculations. Finally, in a rare
bootstrap
realisation, no valid weighted contribution can occur. To make the algorithm stable, the given replicate is treated as
undefined and is not included in the variance estimation.
\subsection*{FCS fiting method}

\indent The module for fitting FCS curves supports four file types. \textit{i)} the \emph{.corr} files that are created by \fcsit\
software as an output of the Import PTU method, \textit{ii)} The three-column, tab-separated datafiles (\emph{.dat}) with columns
marked as X (lagtime column), Y (correlation values $G$), and Y\_err (standard error of $G$), \textit{iii)} The two-column,
tab-separated datafiles (\emph{.dat}) containing columns marked as X (lagtime column), Y (correlation values $G$), and
\textit{iv)} the multicolumn \emph{.dat} files in the format of files exported from the \spt\ software. The imported multicolumn
file is split into single files (one file per correlation curve) and stored in a subfolder named after the original data file. \\
\indent To fit the data, the user can choose between nine predefined fitting models, including: one, two or three-component simple
diffusion models with or without triplets, an anomalous diffusion model, one-component diffusion with
rotation,\cite{Michalski2024} and two-component diffusion where one component performs translational movements and the second
component undergoes the translational and rotational diffusion.\cite{Michalski2024} The users can also define their own models
using a simple text editor as described in the software's manual.\cite{Fcsit_Kalwarczyk2026}\\
\indent The experimental correlation
curves along with the fits are displayed in a standard manner (logscale for the $\tau$ axis and linear scale for $\Gtau{}{}$
axis), and as a log-log plot. The latter is convenient when the correlation curves for long $\tau$ values are highly scattered,
mostly due to low statistics of the autocorrelation function. In such cases, the user can limit the fitting range, which slightly
improves the fit's quality. Additionally, the \dpg\ interface allows for live-mode pre-adjustment of the fitting parameters. The
fitting can be performed for a set of data files, and the results can be exported as parameter tables and plots.
\begin{figure}[t!]
\includegraphics[width=1\linewidth]{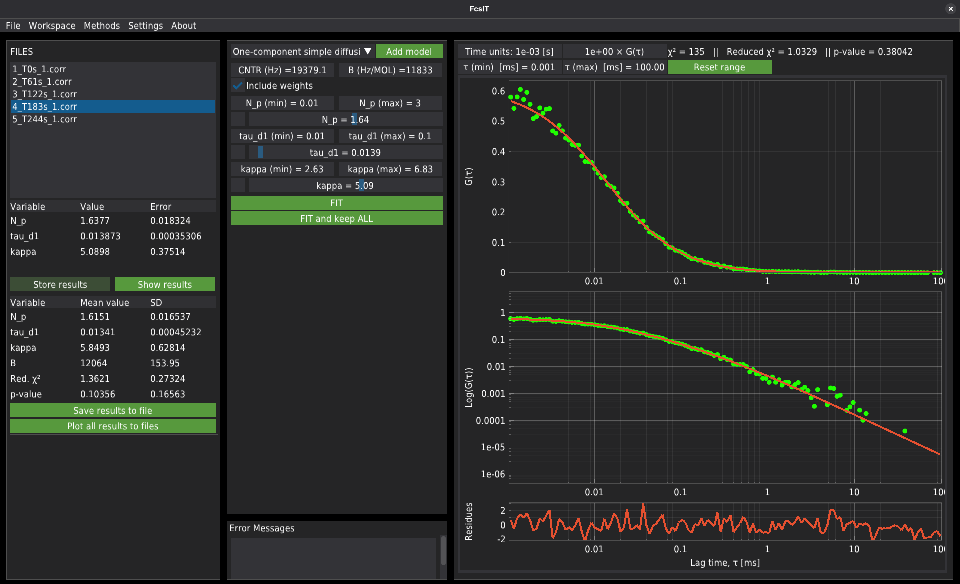}
 \caption{\label{fig:FIT-interface}\textbf{The interface of the FCS fitting module.}
On the left-hand side of the screenshot, there is a list of files under analysis. Below is a table containing fitted values and
errors for the selected data file. The second table contains the mean of the fitted values, averaged over all data files. The
middle panel contains the model selection panel. Below that panel, there is the panel listing the model parameters. This panel is
most useful, as the user can pre-adjust the initial values of the fit manually, limit the fitting ranges for each parameter, and
make the variables fixed or free. Pressing the \textit{FIT} button starts fitting the single data file. To store the data, users
need to add the results manually to the database by pressing the Store results button. The \textit{FIT and keep ALL} button
performs fitting for all data files and automatically adds the fitting results to the database. Each curve is plotted on the
right-hand side of the screen in two ways. The semi-logarithmic plot with the linear scale on the $\Gtau{}{}$ axis and
logarithmic
scale on the $\tau$ axis. Below is a plot on a double-logarithmic scale. At the bottom, residuals are plotted.}
\end{figure}

\section*{Results and Disussion}
\indent The \fcsit\ software and its algorithms were validated against two data sets: simulated data generated with the MCell and
FERNET tools, and experimental data obtained for Rhodamine 110 in water.\\
\begin{figure}[t!]
\includegraphics[width=1\linewidth]{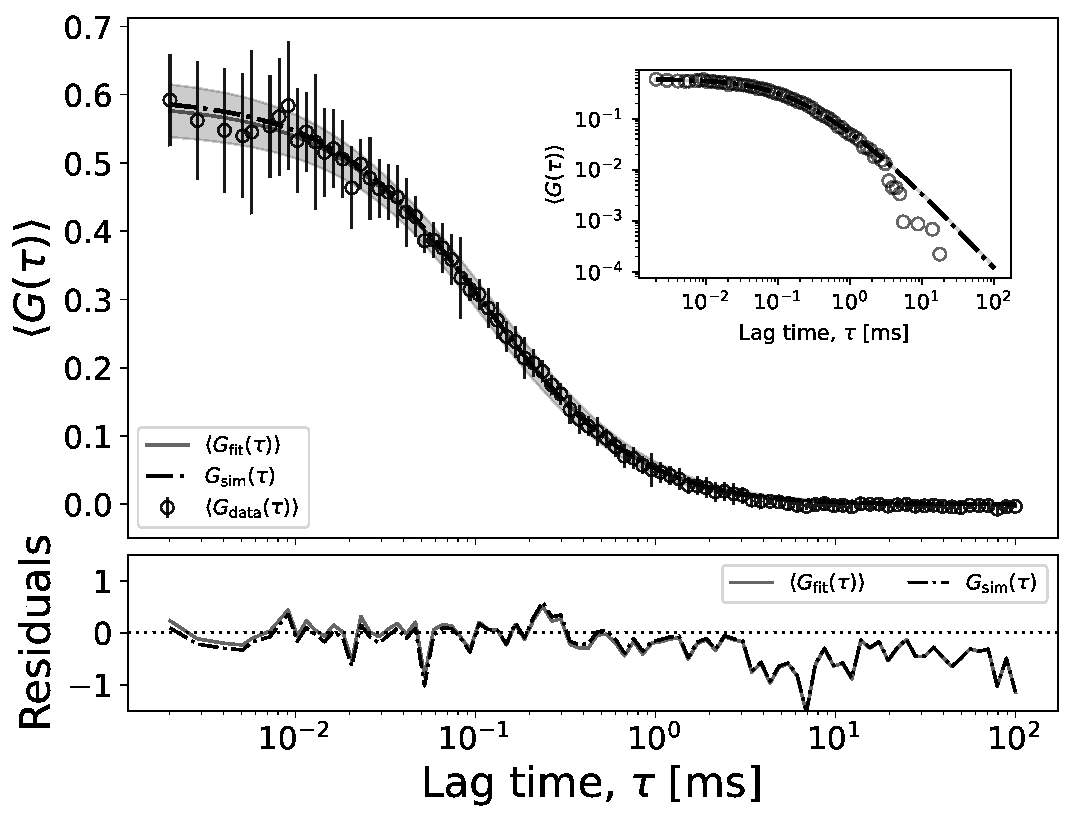}
 \caption{\label{fig:sim_fcs_comp}\textbf{Mean autocorrelation function of simulated data.} The plot shows the mean
data, $\langle G_\mathrm{data}\left(\tau\right)\rangle$, (black
circles) with errorbars corresponding the the standard deviation of the data $\sigma_{G_{\mathrm{data}}}(\tau)$, and the
mean fit, $\langle G_\mathrm{fit}\left(\tau\right)\rangle$, (gray line) averaged over 16 autocorrelation curves obtained
from simulated time traces. The mean data
were compared with the autocorrelation curve (black, dash-dotted line) generated using the $\Np$, $\taudi{}$, and
$\kappa$ parameters as input
values for MCell and FERNET, $G_\mathrm{sim}\left(\tau\right)$. Shaded area corresponds to the standard deviation of the
fit, $\sigma_{G_{\mathrm{fit}}}(\tau)$.}
\end{figure}

\begin{figure}[t!]
\includegraphics[width=1\linewidth]{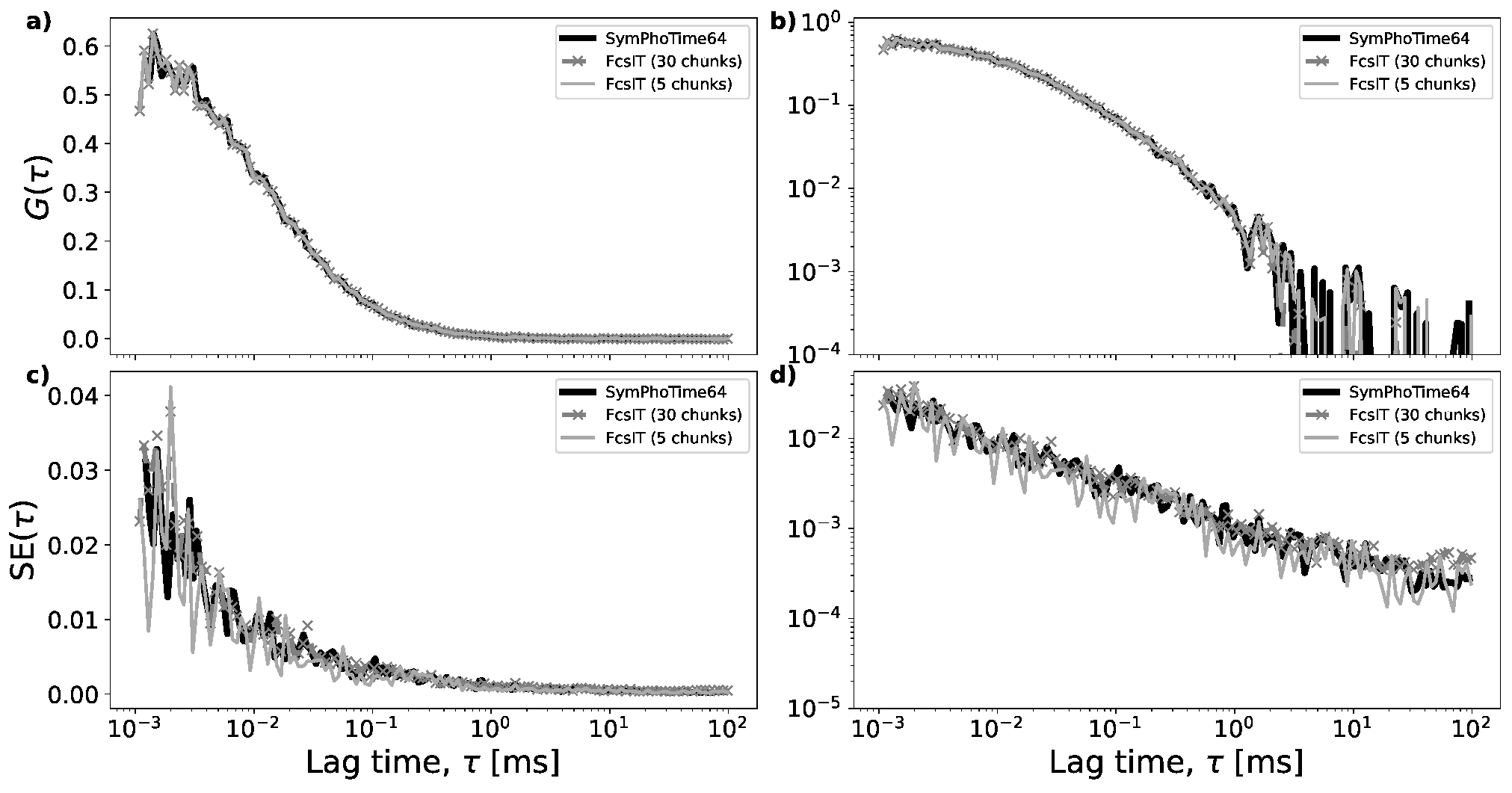}\\
\includegraphics[width=1\linewidth]{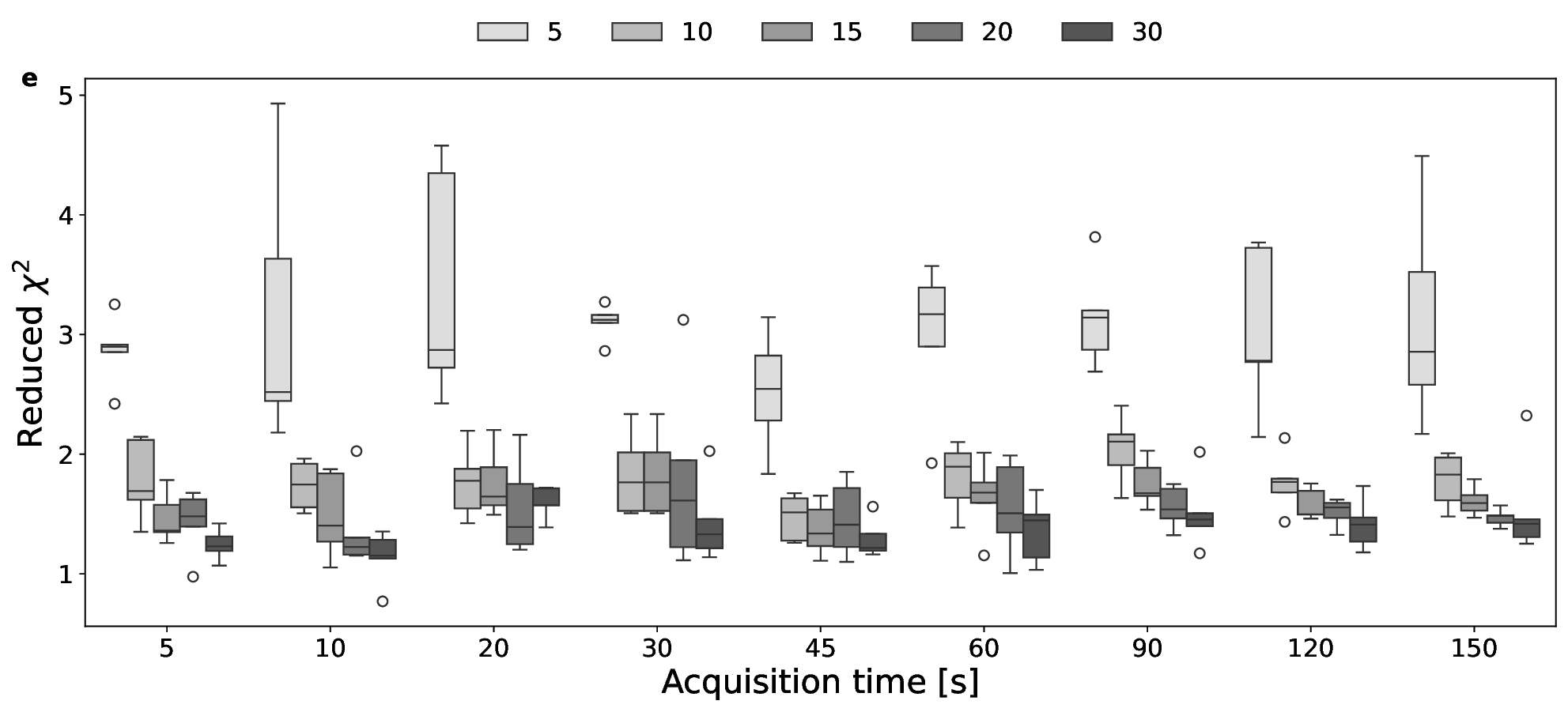}
 \caption{\label{fig:ACF_comp} \textbf{Exemplary autocorrelation curves for rhodamine 110 in water and reduced $\chi^2$ values for
fits of experimental data.} Figures \textbf{a} and \textbf{b} compare autocorrelation curves correlated with the \spt\ and \fcsit\
software. Figure \textbf{a} presents data where the lag time $\tau$ is presented on the logarithmic scale, while $\Gtau{}{}$ is on
a linear scale. Figure \textbf{b} displays the same data using a logarithmic scale for both axes. Figures \textbf{c} and
\textbf{d} compare the standard errors (SE) of the autocorrelation functions calculated with both programs; the data are presented
in a
manner analogous to figures \textbf{a} and \textbf{b}. The SE amplitudes calculated with \spt\ and \fcsit\, using 30 chunks,
are comparable. The SE amplitude for data correlated with \fcsit\ using only 5 chunks is visibly higher, which translates into
higher values of the reduced $\chi^2$ displayed in Figure \textbf{e} and reported in Table \ref{tab:fit_values}.}
\end{figure}

\begin{table}[th!]
\caption{\label{tab:fit_values} \textbf{The mean values of fitting parameters.} In the table, we present the mean parameter values
obtained from fitting the experimental data collected with \spt\ and \fcsit. The data were averaged across 5 curves acquired at
each acquisition time. The data were fitted using a simple diffusion model with only three parameters: the number of particles in
the focal volume ($\Np$), the diffusion time ($\taudi{}$), and the structure parameter ($\kappa$). The reduced $\chi^2$ values
characterise the quality of the fits.}

\begin{center}
\begin{tabular}{l|cccc}
\toprule
 & \multicolumn{4}{r}{SymPhoTime64} \\
 & $N_\mathrm{p}$ & $\tau_\mathrm{d}$ [$\mu$s] & $\kappa$ & Red. $\chi^2$ \\
Acquisition time [s] &  &  &  &  \\
\midrule
5 & 1.66$\pm$0.09 & 13.97$\pm$0.19 & 6.15$\pm$0.51 & 0.93$\pm$0.11 \\
10 & 1.62$\pm$0.03 & 13.21$\pm$0.2 & 6.33$\pm$0.41 & 0.95$\pm$0.12 \\
20 & 1.61$\pm$0.03 & 13.86$\pm$0.13 & 5.32$\pm$0.31 & 1.04$\pm$0.15 \\
30 & 1.6$\pm$0.04 & 13.61$\pm$0.08 & 5.32$\pm$0.13 & 2.51$\pm$0.55 \\
45 & 1.58$\pm$0.03 & 13.34$\pm$0.03 & 5.84$\pm$0.07 & 1.67$\pm$0.32 \\
60 & 1.6$\pm$0.02 & 13.19$\pm$0.03 & 5.83$\pm$0.09 & 1.49$\pm$0.16 \\
90 & 1.58$\pm$0.02 & 13.27$\pm$0.01 & 5.99$\pm$0.04 & 1.36$\pm$0.13 \\
120 & 1.58$\pm$0.02 & 13.19$\pm$0.01 & 5.77$\pm$0.03 & 1.11$\pm$0.08 \\
150 & 1.59$\pm$0.01 & 13.28$\pm$0.02 & 5.82$\pm$0.03 & 1.36$\pm$0.23 \\
\bottomrule
\end{tabular}

\begin{tabular}{l|cccc}
\toprule
 & \multicolumn{4}{r}{FcsIT 5 chunks} \\
 & $N_\mathrm{p}$ & $\tau_\mathrm{d}$ [$\mu$s] & $\kappa$ & Red. $\chi^2$ \\
Acquisition time [s] &  &  &  &  \\
\midrule
5 & 1.71$\pm$0.06 & 14.71$\pm$0.29 & 6.62$\pm$0.89 & 2.87$\pm$0.3 \\
10 & 1.6$\pm$0.05 & 13.22$\pm$0.21 & 6.86$\pm$1.21 & 3.14$\pm$1.14 \\
20 & 1.6$\pm$0.07 & 13.53$\pm$0.2 & 6.06$\pm$0.31 & 3.39$\pm$1.0 \\
30 & 1.6$\pm$0.03 & 13.05$\pm$0.1 & 7.08$\pm$0.31 & 3.1$\pm$0.15 \\
45 & 1.59$\pm$0.05 & 13.35$\pm$0.03 & 6.39$\pm$0.14 & 2.53$\pm$0.5 \\
60 & 1.62$\pm$0.03 & 13.65$\pm$0.04 & 5.74$\pm$0.13 & 2.99$\pm$0.65 \\
90 & 1.61$\pm$0.03 & 13.78$\pm$0.04 & 6.17$\pm$0.16 & 3.14$\pm$0.43 \\
120 & 1.61$\pm$0.02 & 13.57$\pm$0.03 & 5.78$\pm$0.06 & 3.04$\pm$0.7 \\
150 & 1.6$\pm$0.02 & 13.33$\pm$0.03 & 6.38$\pm$0.05 & 3.12$\pm$0.91 \\
\bottomrule
\end{tabular}

\begin{tabular}{l|cccc}
\toprule
 & \multicolumn{4}{r}{FcsIT 30 chunks} \\
 & $N_\mathrm{p}$ & $\tau_\mathrm{d}$ [$\mu$s] & $\kappa$ & Red. $\chi^2$ \\
Acquisition time [s] &  &  &  &  \\
\midrule
5 & 1.71$\pm$0.1 & 15.4$\pm$0.43 & 3.97$\pm$0.44 & 1.24$\pm$0.13 \\
10 & 1.63$\pm$0.03 & 13.62$\pm$0.13 & 5.29$\pm$0.22 & 1.14$\pm$0.23 \\
20 & 1.63$\pm$0.02 & 13.95$\pm$0.08 & 5.0$\pm$0.2 & 1.59$\pm$0.14 \\
30 & 1.61$\pm$0.03 & 13.71$\pm$0.06 & 5.45$\pm$0.13 & 1.43$\pm$0.35 \\
45 & 1.6$\pm$0.04 & 13.5$\pm$0.04 & 5.96$\pm$0.04 & 1.29$\pm$0.16 \\
60 & 1.62$\pm$0.02 & 13.41$\pm$0.03 & 5.85$\pm$0.12 & 1.36$\pm$0.27 \\
90 & 1.59$\pm$0.02 & 13.5$\pm$0.04 & 6.03$\pm$0.06 & 1.51$\pm$0.31 \\
120 & 1.6$\pm$0.02 & 13.38$\pm$0.02 & 5.83$\pm$0.07 & 1.41$\pm$0.21 \\
150 & 1.61$\pm$0.01 & 13.4$\pm$0.03 & 6.07$\pm$0.05 & 1.55$\pm$0.44 \\
\bottomrule
\end{tabular}

\end{center}
\end{table}

\indent The calculated correlations of the simulated data and the corresponding fits are depicted in Figures S1\_0 to S1\_15 from
the
Supporting Information. The individual autocorrelation curves were fitted with the simple diffusion model. The $\Gtau{}{}$ data
for each curve exhibit a high degree of scattering and deviations of the fits from the theoretical curves calculated using the
input parameters for MCell/FERNET simulations ($D$=100 $\mumsq$, $\omega_\mathrm{xy}$=0.212 $\mum$,
$N$=1.68, $\taudi{}$ = 0.112 ms). It is due to numerical noise
in the simulations and to the calculations being performed on short time-trace fragments, about 1 s long. For analysis purposes,
the sixteen simulated autocorrelation curves and their fits were averaged. In Figure \ref{fig:sim_fcs_comp}, the mean
autocorrelation data and the mean fit were compared to the theoretical curve calculated using the MCell/FERNET input parameters.
The mean fit and the theoretical curve describe the mean autocorrelation data in nearly identical fashion.\\
\indent The experimental data correlated with the \fcsit\ software were compared with those correlated with \spt\ from PicoQuant.
Figures \ref{fig:ACF_comp}\textbf{a} and \ref{fig:ACF_comp}\textbf{b} depict exemplary autocorrelation curves represented in a
semi-log scale (Figure \textbf{a}) and a log-log scale (Figure \textbf{b}). For data correlated with the \fcsit\ software, curves
correlated using 5 or 30 chunks were compared. Intuitively, the number of chunks influences the values of the standard errors (SE)
of the autocorrelation curves. The more chunks, the lower the uncertainty of the given $\tau$ value, as reflected in the decrease
in SE amplitude depicted in Figures \ref{fig:ACF_comp}\textbf{c} and \ref{fig:ACF_comp}\textbf{d}. The SE values obtained from the
\spt\ and \fcsit\ programs are shown on the semi-log and log-log scales in Figures \ref{fig:ACF_comp}\textbf{c} and
\ref{fig:ACF_comp}\textbf{d}, respectively.
The number of chunks used in the correlation procedure strongly influences the fit quality (as reflected in the $\chi^2$ value).
To
determine the optimal number of chunks, the autocorrelation curves obtained from the \fcsit\ for 5 different chunk sizes (5, 10,
15, 20, and 30) were compared. The reduced $\chi^2$ values for all tested acquisition times and chunk numbers are shown in Figure
\ref{fig:ACF_comp}\textbf{e}. Based on the results shown in Figure \ref{fig:ACF_comp}\textbf{e}, 30 chunks were set as the default
for calculating correlation curves in \fcsit. However, the users are free to modify this value according to their preferences.\\
\indent Additionally, the fitting results obtained for data correlated with both the \spt\ and \fcsit\ programs (all fits were
performed in \fcsit) were compared. Table \ref{tab:fit_values} shows the mean data averaged over five autocorrelation curves, for
all acquisition times. Results for individual curves used for averaging are presented in Table S1 from the Supporting
Information.
Both programs return similar values of the fitting parameters. A significant difference between the values is observed for the
shortest acquisition time (5s). This observation was expected, as in the standard FCS technique, measurements at such short
acquisition times are at the lower limit of the technique and are subject to substantial statistical uncertainty.\\

\section*{Conclusions}
\indent The \fcsit\ software for calculating the correlation and fitting fluorescence correlation spectroscopy data has been
presented.
The
software can read and correlate TTTR data only from .ptu files created in \spt\ using the fast, efficient asynchronous
TCSPC algorithm.\cite{Wahl2003,Waithe2021} It can also correlate time-binned data using the standard multipletau
method.\cite{Mueller2012Multipletau} Fitting of the correlation curves can be performed for correlation data calculated with
\fcsit\ or stored in ASCII files, making \fcsit\ independent of the FCS setup vendor. The flexible GUI engine, design, and an
intuitive, user-friendly interface allow less experienced users to gain an intuitive understanding of how fitting parameters
influence the shape of correlation curves. This knowledge can lead to a better experimental design. Nine implemented FCS fitting
models, and the flexibility to add user-defined models makes \fcsit\ a powerful analytical tool for the most demanding systems,
including living cells. Finally, the software’s open-source character allows for modifications and further development. Exemplary
templates for building your own extensions are available on the repository site.

% \section*{Acknowledgments}
% We thank dr Karina Kwapiszewska for fruitfull discusions about future development of the tool.

\section*{Data availability}
The data that support the findings of this study are deposited
in RepOD repository (\url{https://repod.icm.edu.pl/}) and will be made publicly available upon publication
of the final article.

\section*{Code availability}
The \fcsit\ code related to this article is available at \url{https://doi.org/10.5281/zenodo.19351494} (version
v1.0.1)\cite{Fcsit_Kalwarczyk2026}
The development repository is hosted on GitHub (\url{https://github.com/TKmist/FcsIT}.)
\section*{Supporting information}
The supporting information for this paper is available as a PDF file containing 15 plots (S1\_0 to S1\_15) and one S1 table.
% \nolinenumbers

%This is where your bibliography is generated. Make sure that your .bib file is actually called library.bib
\bibliography{References}

@Article{Elson1974,
  author    = {Elson, Elliot L. and Magde, Douglas},
  journal   = {Biopolymers},
  title     = {Fluorescence correlation spectroscopy. I. Conceptual basis and theory},
  year      = {1974},
  number    = {1},
  pages     = {1--27},
  volume    = {13},
  doi       = {10.1002/bip.1974.360130102},
  publisher = {Wiley Online Library},
}

@Article{Sengupta2003,
  author    = {Sengupta, Parijat and Garai, K and Balaji, J and Periasamy, N and Maiti, S},
  journal   = {Biophysical journal},
  title     = {Measuring size distribution in highly heterogeneous systems with fluorescence correlation spectroscopy},
  year      = {2003},
  number    = {3},
  pages     = {1977--1984},
  volume    = {84},
  publisher = {Elsevier},
}

@misc{Mueller2012Multipletau,
  author       = {Paul Müller},
  title        = {Python multiple-tau algorithm},
  year         = {2012},
  note         = {Version x.x.x},
  howpublished = {\url{https://pypi.python.org/pypi/multipletau/}},
  urldate      = {YYYY-MM-DD}
}

@Misc{Krieger2015,
  author       = {Jan Wolfgang Krieger and Jörg Langowski},
  howpublished = {[web page] \verb+http://www.dkfz.de/Macromol/quickfit/+},
  title        = {QuickFit 3.0: A data evaluation application for biophysics},
  year         = {2015},
}

@Article{Mueller2014,
  author  = {M{\"u}ller, Paul and Koberling, Felix and Enderlein, J{\"o}rg},
  journal = {Bioinformatics},
  title   = {PyCorrFit—generic data evaluation for fluorescence correlation spectroscopy},
  year    = {2014},
  number  = {17},
  pages   = {2532--2533},
  volume  = {30},
  doi     = {10.1093/bioinformatics/btu321},
}

@Article{Ries2010,
  author    = {Ries, Jonas and Bayer, Mathias and Cs{\'u}cs, G{\'a}bor and Dirkx, Ronald and Solimena, Michele and Ewers, Helge and Schwille, Petra},
  journal   = {Optics express},
  title     = {Automated suppression of sample-related artifacts in Fluorescence Correlation Spectroscopy},
  year      = {2010},
  number    = {11},
  pages     = {11073--11082},
  volume    = {18},
  publisher = {Optical Society of America},
}

@Article{Waithe2021,
  author    = {Waithe, Dominic},
  journal   = {Nature Photonics},
  title     = {Open-source browser-based software simplifies fluorescence correlation spectroscopy data analysis},
  year      = {2021},
  number    = {11},
  pages     = {790--791},
  volume    = {15},
  publisher = {Nature Publishing Group UK London},
}

@Misc{Ingargiola,
  author       = {Antonino Ingargiola},
  howpublished = {\url{https://github.com/paulmueller/pycorrelate}},
  title        = {PyCorrelate - fast and accurate cross-correlation for photon timestamp data},
  doi          = {10.5281/zenodo.2653505},
}

@Article{Peulen2025,
  author   = {Peulen, Thomas-Otavio and Hemmen, Katherina and Greife, Annemarie and Webb, Benjamin M and Felekyan, Suren and Sali, Andrej and Seidel, Claus A M and Sanabria, Hugo and Heinze, Katrin G},
  journal  = {Bioinformatics},
  title    = {tttrlib: modular software for integrating fluorescence spectroscopy, imaging, and molecular modeling},
  year     = {2025},
  issn     = {1367-4811},
  month    = {01},
  number   = {2},
  pages    = {btaf025},
  volume   = {41},
  doi      = {10.1093/bioinformatics/btaf025},
  eprint   = {https://academic.oup.com/bioinformatics/article-pdf/41/2/btaf025/61530906/btaf025.pdf},
  url      = {https://doi.org/10.1093/bioinformatics/btaf025},
}

@Article{Priest2019,
  author    = {Priest, David G and Solano, Ashleigh and Lou, Jieqiong and Hinde, Elizabeth},
  journal   = {Biochemical Society Transactions},
  title     = {Fluorescence fluctuation spectroscopy: an invaluable microscopy tool for uncovering the biophysical rules for navigating the nuclear landscape},
  year      = {2019},
  number    = {4},
  pages     = {1117--1129},
  volume    = {47},
  publisher = {Portland Press Ltd.},
}

@Article{Shang2017,
  author    = {Shang, Li and Nienhaus, G Ulrich},
  journal   = {Accounts of chemical research},
  title     = {In situ characterization of protein adsorption onto nanoparticles by fluorescence correlation spectroscopy},
  year      = {2017},
  number    = {2},
  pages     = {387--395},
  volume    = {50},
  publisher = {ACS Publications},
}

@Article{Klapper2015,
  author    = {Klapper, Yvonne and Maffre, Pauline and Shang, Li and Ekdahl, Kristina N and Nilsson, Bo and Hettler, Simon and Dries, Manuel and Gerthsen, Dagmar and Nienhaus, G Ulrich},
  journal   = {Nanoscale},
  title     = {Low affinity binding of plasma proteins to lipid-coated quantum dots as observed by in situ fluorescence correlation spectroscopy},
  year      = {2015},
  number    = {22},
  pages     = {9980--9984},
  volume    = {7},
  publisher = {Royal Society of Chemistry},
}

@Article{Ries2009,
  author    = {Ries, Jonas and Yu, Shuizi Rachel and Burkhardt, Markus and Brand, Michael and Schwille, Petra},
  journal   = {Nature methods},
  title     = {Modular scanning FCS quantifies receptor-ligand interactions in living multicellular organisms},
  year      = {2009},
  number    = {9},
  pages     = {643--645},
  volume    = {6},
  publisher = {Nature Publishing Group US New York},
}

@Article{Michalska2018,
  author    = {Michalska, Bernadeta Maria and Kwapiszewska, Karina and Szczepanowska, Joanna and Kalwarczyk, Tomasz and Patalas-Krawczyk, Paulina and Szczepa{\'n}ski, Krzysztof and Ho{\l}yst, Robert and Duszy{\'n}ski, Jerzy and Szyma{\'n}ski, J{\k{e}}drzej},
  journal   = {Scientific reports},
  title     = {Insight into the fission mechanism by quantitative characterization of Drp1 protein distribution in the living cell},
  year      = {2018},
  number    = {1},
  pages     = {8122},
  volume    = {8},
  publisher = {Nature Publishing Group UK London},
}

@Article{Kwapiszewska2019,
  author    = {Kwapiszewska, Karina and Kalwarczyk, Tomasz and Michalska, Bernadeta and Szczepa{\'n}ski, Krzysztof and Szyma{\'n}ski, J{\k{e}}drzej and Patalas-Krawczyk, Paulina and Andryszewski, Tomasz and Iwan, Michalina and Duszy{\'n}ski, Jerzy and Ho{\l}yst, Robert},
  journal   = {Scientific reports},
  title     = {Determination of oligomerization state of Drp1 protein in living cells at nanomolar concentrations},
  year      = {2019},
  number    = {1},
  pages     = {5906},
  volume    = {9},
  publisher = {Nature Publishing Group UK London},
}

@Article{Kwapiszewska2020,
  author    = {Kwapiszewska, Karina and Szczepa{\'n}ski, Krzysztof and Kalwarczyk, Tomasz and Michalska, Bernadeta and Patalas-Krawczyk, Paulina and Szyma{\'n}ski, J{\k{e}}drzej and Andryszewski, Tomasz and Iwan, Michalina and Duszy{\'n}ski, Jerzy and Ho{\l}yst, Robert},
  journal   = {The journal of physical chemistry letters},
  title     = {Nanoscale viscosity of cytoplasm is conserved in human cell lines},
  year      = {2020},
  number    = {16},
  pages     = {6914--6920},
  volume    = {11},
  publisher = {ACS Publications},
}

@Article{Holyst2009,
  author    = {Holyst, Robert and Bielejewska, Anna and Szyma{\'n}ski, J{\k{e}}drzej and Wilk, Agnieszka and Patkowski, Adam and Gapi{\'n}ski, Jacek and {\.Z}ywoci{\'n}ski, Andrzej and Kalwarczyk, Tomasz and Kalwarczyk, Ewelina and Tabaka, Marcin and others},
  journal   = {Physical Chemistry Chemical Physics},
  title     = {Scaling form of viscosity at all length-scales in poly (ethylene glycol) solutions studied by fluorescence correlation spectroscopy and capillary electrophoresis},
  year      = {2009},
  number    = {40},
  pages     = {9025--9032},
  volume    = {11},
  publisher = {Royal Society of Chemistry},
}

@Article{Kalwarczyk2011,
  author    = {Kalwarczyk, Tomasz and Ziebacz, Natalia and Bielejewska, Anna and Zaboklicka, Ewa and Koynov, Kaloian and Szymanski, Jedrzej and Wilk, Agnieszka and Patkowski, Adam and Gapinski, Jacek and Butt, Hans-J{\"u }rgen and others},
  journal   = {Nano letters},
  title     = {Comparative analysis of viscosity of complex liquids and cytoplasm of mammalian cells at the nanoscale},
  year      = {2011},
  number    = {5},
  pages     = {2157--2163},
  volume    = {11},
  publisher = {ACS Publications},
}

@Article{Kapusta2007,
  author    = {Kapusta, Peter and Wahl, Michael and Benda, Ale{\v{s}} and Hof, Martin and Enderlein, J{\"o}rg},
  journal   = {Journal of fluorescence},
  title     = {Fluorescence lifetime correlation spectroscopy},
  year      = {2007},
  pages     = {43--48},
  volume    = {17},
  publisher = {Springer},
}

@Article{Enderlein2005,
  author    = {Enderlein, J{\"o}rg and Gregor, Ingo},
  journal   = {Review of scientific instruments},
  title     = {Using fluorescence lifetime for discriminating detector afterpulsing in fluorescence-correlation spectroscopy},
  year      = {2005},
  number    = {3},
  volume    = {76},
  publisher = {AIP Publishing},
}

@Article{Wahl2003,
  author    = {Wahl, Michael and Gregor, Ingo and Patting, Matthias and Enderlein, J{\"o}rg},
  journal   = {Optics express},
  title     = {Fast calculation of fluorescence correlation data with asynchronous time-correlated single-photon counting},
  year      = {2003},
  number    = {26},
  pages     = {3583--3591},
  volume    = {11},
  publisher = {Optical Society of America},
}

@Article{Rich2013,
  author    = {Rich, Ryan M and Stankowska, Dorota L and Maliwal, Badri P and S{\o}rensen, Thomas Just and Laursen, Bo W and Krishnamoorthy, Raghu R and Gryczynski, Zygmunt and Borejdo, Julian and Gryczynski, Ignacy and Fudala, Rafal},
  journal   = {Analytical and bioanalytical chemistry},
  title     = {Elimination of autofluorescence background from fluorescence tissue images by use of time-gated detection and the AzaDiOxaTriAngulenium (ADOTA) fluorophore},
  year      = {2013},
  number    = {6},
  pages     = {2065--2075},
  volume    = {405},
  publisher = {Springer},
}

@Article{Kondo2025,
  author    = {Kondo, Ryoichi and Yamamoto, Akio and Endo, Tomohiro},
  journal   = {Journal of Nuclear Science and Technology},
  title     = {Sub-pin level distribution tallies and statistical error estimation with POD tallies in two-dimensional C5G7 benchmark},
  year      = {2025},
  pages     = {1--12},
  publisher = {Taylor \& Francis},
}

@Article{Wohland2001,
  author    = {Wohland, Thorsten and Rigler, Rudolf and Vogel, Horst},
  journal   = {Biophysical journal},
  title     = {The standard deviation in fluorescence correlation spectroscopy},
  year      = {2001},
  number    = {6},
  pages     = {2987--2999},
  volume    = {80},
  publisher = {Elsevier},
}

@Article{Haerdle2003,
  author    = {H{\"a}rdle, Wolfgang and Horowitz, Joel and Kreiss, Jens-Peter},
  journal   = {International Statistical Review},
  title     = {Bootstrap methods for time series},
  year      = {2003},
  number    = {2},
  pages     = {435--459},
  volume    = {71},
  publisher = {Wiley Online Library},
}

@Book{Politis1991,
  author    = {Politis, Dimitris N and Romano, Joseph P},
  publisher = {Purdue University. Department of Statistics},
  title     = {A circular block-resampling procedure for stationary data},
  year      = {1991},
}

@Article{Koppel1974,
  author    = {Koppel, Dennis E},
  journal   = {Physical Review A},
  title     = {Statistical accuracy in fluorescence correlation spectroscopy},
  year      = {1974},
  number    = {6},
  pages     = {1938},
  volume    = {10},
  publisher = {APS},
}

@Book{Davison1997,
  author    = {A. C. Davison and D. V. Hinkley},
  publisher = {Cambridge university press},
  title     = {Bootstrap methods and their application},
  year      = {1997},
  number    = {1},
  pages     = {405-408},
}

@Article{Michalski2024,
  author    = {Michalski, Jaros{\l}aw and Kalwarczyk, Tomasz and Kwapiszewska, Karina and Enderlein, J{\"o}rg and Poniewierski, Andrzej and Karpi{\'n}ska, Aneta and Kucharska, Karolina and Ho{\l}yst, Robert},
  journal   = {Soft Matter},
  title     = {Rotational and translational diffusion of biomolecules in complex liquids and HeLa cells},
  year      = {2024},
  number    = {29},
  pages     = {5810--5821},
  volume    = {20},
  publisher = {Royal Society of Chemistry},
}

@Article{Stiles1996,
  author  = {Stiles, Joel R and Van Helden, Dirk and Bartol Jr, TM and Salpeter, Edwin E and Salpeter, Miriam M},
  journal = {Proceedings of the National Academy of Sciences},
  title   = {Miniature endplate current rise times less than 100 microseconds from improved dual recordings can be modeled with passive acetylcholine diffusion from a synaptic vesicle.},
  year    = {1996},
  number  = {12},
  pages   = {5747--5752},
  volume  = {93},
}

@Article{Angiolini2015,
  author    = {Angiolini, Juan and Plachta, Nicolas and Mocskos, Esteban and Levi, Valeria},
  journal   = {Biophysical Journal},
  title     = {Exploring the dynamics of cell processes through simulations of fluorescence microscopy experiments},
  year      = {2015},
  number    = {11},
  pages     = {2613--2618},
  volume    = {108},
  publisher = {Elsevier},
}

@Article{Kerr2008,
  author    = {Kerr, Rex A and Bartol, Thomas M and Kaminsky, Boris and Dittrich, Markus and Chang, Jen-Chien Jack and Baden, Scott B and Sejnowski, Terrence J and Stiles, Joel R},
  journal   = {SIAM journal on scientific computing},
  title     = {Fast Monte Carlo simulation methods for biological reaction-diffusion systems in solution and on surfaces},
  year      = {2008},
  number    = {6},
  pages     = {3126--3149},
  volume    = {30},
  publisher = {SIAM},
}

@Misc{Fcsit_Kalwarczyk2026,
  author    = {Kalwarczyk, Tomasz},
  note      = {Version 1.0.1, Zenodo, Software, DOI:10.5281/zenodo.19351493. (The development repository is hosted on GitHub: \url{https://github.com/TKmist/FcsIT}. The step by step manual is available at the repository site \url{https://tkmist.github.io/FcsIT/})},
  title     = {FcsIT - The simple and easy to use tool for fitting and correlation of the Fluorescence Correlation Spectroscopy data.},
  year      = {2026},
  doi       = {10.5281/zenodo.19351493},
  publisher = {Zenodo},
  version   = {1.0.1},
}

@Misc{Rohilla2019,
  author = {Rohilla, Sumeet},
  note   = {The development repository is hosted on GitHub: \url{https://github.com/SumeetRohilla/readPTU\_FLIM}. For the puroses of the FcsIT software we used modified version of the repository from the branch entitled "NIKON\_correction" available at \url{https://github.com/TKmist/readPTU\_FLIM/NIKON\_correction}},
  title  = {readPTU\_FLIM},
  year   = {2019},
}

%This defines the bibliographies style. Search online for a list of available styles.
\bibliographystyle{unsrtnat}

\end{document}